\documentstyle[12pt,epsf]{article}

\title{EPR effect in gravitational field: \\
nature of non-locality\thanks{Published in Phys. Lett. A269, 204-208 (2000)}}
\author{
Horst von Borzeszkowski$^{{a}}$\thanks{Email: 
borzeszk@itp.physik.tu-berlin.de}
\and
Michael B. Mensky$^{{a}{b}}$\thanks{Email: mensky@sci.lebedev.ru}
\\[3pt]$^{{a}}$
Institut f\"ur Theoretische Physik,
Technische Universit\"at Berlin\\
Hardenbergstr. 36,
D-10623 Berlin, Germany\\
\\[10pt]$^{{b}}$
P.N.Lebedev Physical Institute, 53 Leninskii prosp,\\ 
117924 Moscow, Russia\\
}
\date{February 4, 2000}

\newcommand{\marke}{\label}

\newcommand{\eq}[1]{(\ref{#1})}

\newcommand{\eqss}[3]{(\ref{#1},~\ref{#2},~\ref{#3})}
\newcommand{\Eq}{Eq.~\eq}

\newcommand{\Eqss}{Eqs.~\eqss}
\newcommand{\Sect}[1]{Sect.~\ref{#1}}

\newcommand{\partderiv}[2]{\frac{\partial #1}{\partial #2}}

\newcommand{\be}{\begin{equation}}
\newcommand{\ee}{\end{equation}}
\newcommand{\ba}{\begin{eqnarray}}
\newcommand{\ea}{\end{eqnarray}}
\newcommand{\ban}{\begin{eqnarray*}}
\newcommand{\ean}{\end{eqnarray*}}

\newcommand{\ra}{\rangle}

\newcommand{\Aone}{{A$_1$}}
\newcommand{\Atwo}{{A$_2$}}

\begin{document}
\maketitle

\begin{abstract}
The realization of the Einstein-Podolsky-Rosen effect by the correlation 
of spin projections of two particles created in the decay of a single 
scalar particle is considered for particles propagating in 
gravitational field. The absence of a global definition of spatial 
directions makes it unclear whether the correlation may exist in this 
case and, if yes, what directions in distant regions must be 
correlated. It is shown that in a gravitational field an approximate 
correlation may exist and the correlated directions are connected with 
each other by the parallel transport along the world lines of the 
particles. The reason for this is that the actual origin of the quantum 
non-locality is founded in local processes. \end{abstract}

\section{Introduction}\marke{intro}

The Einstein-Podolsky-Rosen (EPR) effect (paradox) 
\cite{EinstPodolRosen35} demonstrates, as is commonly believed, a 
sort of quantum non-locality because the measurement of one of the 
two spatially separated systems changes the state of the other 
provided that they are in an entangled state. This can be described 
as a quantum 
correlation of the systems. We shall clarify the nature of this 
non-locality by considering the EPR effect for spinning particles 
propagating in an external gravitational field. It is not evident that 
the correlation must exist in this case 
because there is no `natural' correspondence between the directions in 
the two spatially separated regions in a curved space-time. We shall 
however see that an approximate correlation may exist even in this case 
and that the correlated directions are connected with each other  
by the parallel transport along the world lines of the 
particles.\footnote{In quantum mechanics the concept of a world line 
may be introduced in the framework of an approximation, see later.} 
This proves that the apparent non-locality is in fact of a local 
nature. 

It is well known that the EPR effect may be realized with the help of 
two spin 1/2 particles forming in the decay of a scalar particle and 
therefore having correlated spin projections. In this case the 
measurement of the spin projection of the first particle resulting say 
in $m_z^{(1)}=1/2$ (spin up) makes the spin projection of the other 
particle definite, $m_z^{(2)}=-1/2$ (spin down), which may be verified 
by its measurement.

This phenomenon is well understood and easily explained by the
consideration of the state of both particles after the decay:
\be
\frac{1}{\sqrt{2}}
%( |+\ra_1 |-\ra_2 -|-\ra_1 |+\ra_2 ).
\left( |\uparrow\ra_1 |\downarrow\ra_2 
-|\downarrow\ra_1 |\uparrow\ra_2 \right).
\marke{entanglState}\ee
This so called entangled state means that the projections of the spins 
of both particles on the same axis $z$ are correlated: the measurement 
of them always gives opposite results. After some period of free 
evolution the particles may have moved far away from each other, but 
the spin state of both particle and therefore the correlation of the 
spin projections will be the same. 

The localization of the particles in two spatially separated regions 
(the key point for the EPR paradox) is no obstacle for a correct 
formulation of such a correlation `in ordinary conditions'. Indeed, 
the statement ``two particles located in different space regions have 
opposite spin projections on axis $z$'' has quite a definite meaning 
under the condition that the gravitational field acting on the 
particles is negligible.  

The situation however radically changes if the particles move in a 
substantial gravitational field. In this case the statement ``If 
$m_z^{(1)}=1/2$, then $m_z^{(2)}=-1/2$" has no sense for spatially 
separated particles, because, in the absence of teleparallelism, the 
axis $z$ in the region where the second particle is located may be 
chosen independently of the $z$ axis in the vicinity of the first 
particle. 

The following questions naturally arise: 1)~Does a quantum 
correlation maintain between the spins of two particles when the 
particles move in a gravitational field and 2)~if it exists, 
how may this correlation be specified (what directions in the two 
distant regions are correlated)? We shall consider these questions 
first on the intuitive level and then in the framework of the 
path-integral description of a spinning particle in an external 
gravitational field (in a curved space-time). The conclusion will be 
that the correlation is in this case approximate and has to be 
formulated with the help of the parallel transport along the 
trajectories (world lines) of the correlated particles.

\section{Arguments of plausibility}\marke{SectPlaus}

Let in a space-time point O a spin-0 particle decay forming two 
particles of spin 1/2, which then propagate to the space-time points 
A$_1$, A$_2$. If this takes place in the flat (Minkowski) space-time 
(see Fig.~\ref{FigEprGrav} left), then, because of the 
teleparallelism, a single, globally defined reference frame may be 
defined in the space-time. The axis $z$ chosen in point A$_1$ 
may correctly be used in an arbitrary point in the 
whole space-time. Therefore, no question arises as to what is the 
$z$-direction in the point A$_2$. The correlation between the two 
particles may be described in terms of this common $z$-direction as 
it has been made in Introduction. 

If however the space-time is curved (Fig.~\ref{FigEprGrav} right), 
then the $z$-direction defined in a certain manner in point A$_1$ does 
not determine any $z$-direction in other points, among them in point 
A$_2$. 
\begin{figure}[ht]
%%Begin InstantTeX Picture
\let\picnaturalsize=N
\def\picsize{4.0in}
\def\picfilename{epr-grav.eps}
%If you do not have the picture file add:
%\let\nopictures=Y
%to the beginning of the file.
\ifx\nopictures Y\else{\ifx\epsfloaded Y\else\input epsf \fi
\let\epsfloaded=Y
\centerline{\ifx\picnaturalsize N\epsfxsize \picsize\fi 
\epsfbox{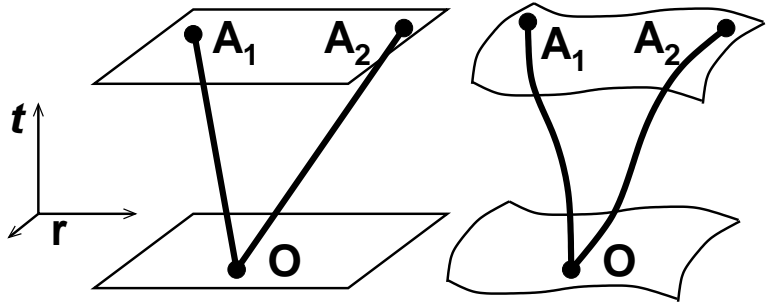}}}\fi
%%End InstantTeX Picture
\caption{Two spinning particles propagating in the flat (left) or 
curved (right) space-time. In the latter case there is no globally 
defined notion of $z$-direction. Instead one has to use local frames. 
However the natural (in respect to the given particles) correspondence 
between the local frames in the points A$_1$ and A$_2$ may be 
established by the parallel transport along the curve A$_1$OA$_2$ 
consisting of the world lines of the particles.}
\marke{FigEprGrav}
\end{figure}

The reason for this difficulty is that there is no global notion of 
the $z$-direction or more generally of spacial directions or space-time 
directions. Such notions may be introduced in an arbitrary (space-time) 
point A with the help of a local frame $n$, which is a basis of the 
tangent space to the curved space-time in the given point. However 
local frames in different points may be chosen independently of each 
other, so there is no natural correspondence between the local frame 
$n_1$ in the point A$_1$ and the local frame $n_2$ in the point A$_2$. 

The correlation between the spin projections in the points A$_1$ and 
A$_2$ can be formulated only if some unambiguous correspondence is 
established between local frames $n_1$ and $n_2$ in the points A$_1$ 
and A$_2$ in such a way that the choice of one of them determines the 
other. This correspondence must be based on some essential features 
of the physical process under consideration. 

If a certain curve connecting the points A$_1$ and A$_2$ is chosen, 
then the correspondence between $n_1$ and $n_2$ may be established 
with the help of the operation of parallel transport along this 
curve. It is quite evident that the only relevant curve for 
this aim is the curve A$_1$OA$_2$ composed of the world lines of the 
particles, so it is natural to choose this curve for the parallel 
transport of the local frames. 

Thus, it seems plausible that the EPR correlation in a gravitational 
field is the following. Choose an arbitrary local frame $n_1$ in 
point A$_1$ and define the 
spin projection of the first particle with respect to the $z$-axis of 
this local frame. Then perform the parallel transport of this local 
frame to the point O along the curve A$_1$O (the world line of the 
first particle but passed backwardly in time) and then along the world 
line OA$_2$ of the second particle to the point A$_2$. A local frame 
$n_2$ in the point A$_2$ results. Define a spin projection of the second 
particle with respect to the $z$-axis of this frame. The correlation 
formulated in Introduction should exist between the thus defined spin 
projections. 

This scheme seems plausible, but of course it must be proved in the 
framework of the dynamical consideration. Particularly, it has to be 
justified 1)~that the operation of parallel transport must be 
taken to identify $z$-directions in the locations of the two particles 
and 2)~that the curve for this parallel transport must be composed of 
the world lines of the particles. Moreover, in the plausible 
formulation we used the purely classical notion of a world line of a 
particle which may only approximately be applied in the case of 
quantum particles. 
These points will be made more precise in the next section 
in the course of the dynamical consideration on the basis of the path- 
integral approach. It will be shown that the above-mentioned 
correlation may actually arise for particles in a gravitational field, 
but only approximately, if the particles are localized around the 
points O, \Aone, \Atwo{} in regions which are narrow in comparison 
with the distances O\Aone{} and O\Atwo. In the general case the 
correlation of spin projections emerging in the decay is destroyed 
during the propagation of the particles in the gravitational field. 

\section{Dynamical proof}\marke{SectDynamic}

In this section we shall shortly introduce a formalism for the 
description of a relativistic spinning particle moving in a curved 
space-time. Then this formalism will be applied to analyze EPR 
corralations in a curved space-time.  

\subsection{The evolution of a relativistic spinning particle}

Let $\psi(x)$ (with $x$ being a space-time point) is an arbitrary solution 
of the (covariant) Klein-Gordon (KG) equation and $U(x,x')$ an 
arbitrary Green function of this equation: 
\ba
(\Box + m^2)\psi(x)&=&0, \marke{KleinGordEq} \\
(\Box + m^2)\,U(x,x')&=&-i\delta(x,x'). \marke{GreenKleinGordEq}
\ea
Then for an arbitrary four-dimensional region $\Omega$ having a 
characteristic function $\eta_\Omega (x)$ (the function which is 
equal to unity in $\Omega$ and zero outside) the following relation 
holds:
\be
-i\int_{\partial\Omega} 
\sigma^{\mu'}\, U(x,x') \stackrel{\mbox{$\leftrightarrow$}}{\nabla}_{\mu'}
\psi(x') = \eta_\Omega (x)\psi(x).
\marke{closedSurf}\ee
Here the three-dimensional integration over the boundary 
$\partial\Omega$ of $\Omega$ is performed with the measure determined 
by the three-form 
\be
\sigma_\mu=\frac{1}{6} \sqrt{-g(x)}\,
\epsilon_{\mu\nu\sigma\lambda}\;
dx^\nu \wedge dx^\sigma \wedge dx^\lambda, 
\marke{sigma}\ee
${\nabla}_{\mu}$ is a covariant derivative and 
$\stackrel{\mbox{$\leftrightarrow$}}{\nabla}_{\mu}$ denotes
$$
\varphi(x)
\stackrel{\mbox{$\leftrightarrow$}}{\nabla}_{\mu}
\psi(x)
= \varphi(x){\nabla}_{\mu}\psi(x)
- ({\nabla}_{\mu}\varphi(x))\psi(x).
$$
The relation \eq{closedSurf} may be derived from \Eq{GreenKleinGordEq}. 

Let $\Omega$ be a 4-dimensional region containing the point $x$ 
and restricted by two space-like surfaces, one in the future and 
the other in the past of $x$. Then the integral \eq{closedSurf} 
is performed over these two surfaces. However for a positive-frequency 
solution of the KG equation (describing a particle but having no 
counterpart related to an antiparticle)\footnote{The 
positive-frequency function is defined in the region of interest 
(where the particle propagates) if a time-like Killing vector field 
exists in this region.} and for $U(x,x')$ being the causal propagator 
the integral over the future surface is zero. Therefore, 
only the integral over the past surface $S$ 
survives in this case (this may be interpreted as the property of 
positive frequencies to propagate only forward in time):
\be
\psi^{(+)}(x) = 
i\int_{S<x} 
\sigma^{\mu'}\, U(x,x') \stackrel{\mbox{$\leftrightarrow$}}{\nabla}_{\mu'}
\psi^{(+)}(x'). 
\marke{relatEvolut}\ee

This is the evolution law for a relativistic particle which may have 
non-zero spin. To analyze the problem of correlation we need to 
investigate properties of the propagator $U(x,x')$. We shall do this 
with the help of the path-integral presentation of this propagator. 

\subsection{The path integral for the propagator}

Different and not always equivalent definitions of the path integral 
in a curved space-time may be given \cite{DeWitt,TMF74,Kleinert}. We 
shall use the definition of \cite{TMF74} which is equivalent to that 
one in \cite{Kleinert}. 

It turns out that it is much simpler to give a path-integral 
representation not for the causal propagator $U(x,x')$ (which is a 
2-point function) but for the corresponding integral operator $U$ 
defined as follows: 
\be
(U\varphi)(x)=\int d^4x\sqrt{-g(x)}\, U(x,x')\,\varphi(x').
\marke{PropagOper}\ee
The path-integral definition of this operator may be given in the  
following way \cite{TMF74}. First the operator $U$ may be expressed in 
terms of the operator $U_{\tau}$ depending on `proper time' $\tau$ 
(which is sometimes called historical time or simply fifth parameter 
\cite{PropTime}): 
\be
U=\int_0^{\infty} d\tau\, e^{-im^2 \tau}U_{\tau}. 
\marke{evolPropTime}\ee
The operator $U_{\tau}$ may be expressed in the form of a path 
integral:  
\be
U_{\tau}=\int d[\xi]\, e^{(-i/4)\int_0^{\tau}d\tau\,
\dot{\xi}^{\alpha}\dot{\xi}_{\alpha}} \, D[\xi]. 
\marke{evolPathInt}\ee
Here the integration is carried out over all paths $[\xi]$ in Minkowski 
space. Paths may be defined as elements of the group of paths (classes 
of continuous curves) \cite{PathGr} or simply as continuous curves 
having the fixed initial point (say, the origin of Minkowski space) 
and an arbitrary final point. The covariant displacement operator 
$D[\xi]$ is defined as an ordered exponential of the integral along 
$[\xi]$: 
\be
D[\xi]= P\exp\left( \int_{[\xi]} d\xi^\alpha\, \nabla_\alpha\right).
\marke{DisplaceOper}\ee
This is in fact the operator of parallel transport along 
those curves in the curved space-time which have the curve $[\xi]$ (in 
Minkowski space) as their evolvent, see \cite{TMF74,EquivPrinc}. 
The integral in \Eq{DisplaceOper} 
includes the operator of the covariant derivative but carrying the 
Lorentzian index $\alpha$ instead of the world index $\mu$ :  
\be
(\nabla_\alpha\varphi)(x)=n_\alpha^\mu(x)\, (\nabla_\mu \varphi)(x). 
\marke{covarDerivStandard}\ee
This formula makes use of a field $n(x)=\{ n_\alpha^\mu(x)\}$ of local 
frames (a section of the fiber bundle of orthonormal local frames) 
which may be chosen arbitrarily and then used in this and the 
subsequent formulas. For the spinning particle corresponding to the 
representation $D$ of the Lorentz group, the covariant derivative 
$\nabla_\mu$ with the world index $\mu$ is defined as 
follows\footnote{We make use here of the conventional definition of 
the covariant derivative instead of the equivalent definition in terms 
of the fiber bundle of local frames which has been used in 
\cite{TMF74}}: 
\be
(\nabla_\mu\,\psi)(x)
=\left[ \partderiv{}{x^\mu}+\widetilde{D}(M_\mu(x)) \right]\psi(x)
\marke{covarDeriv}\ee
where $\widetilde{D}$ is the representation of the Lorentzian Lee 
algebra corresponding to the representation $D$ of the Lorentz group. 
The matrix $M(x)$ belonging to the Lorentzian Lee algebra is 
$$
\left[ M_\mu(x) \right]^\gamma_\beta 
= \left(n^{-1} \right)^\gamma_\nu 
\left[ \partderiv{n^\nu_\beta(x)}{x^\mu}
+n^\sigma_\beta\Gamma^\nu_{\mu\sigma}(x) \right].
$$

The `propagator depending on the proper time' $U_\tau(x,x')$ 
satisfies the relativistic Schr\"odinger equation:
\be
(\Box -i\partderiv{}{\tau})\,U_\tau(x,x')=0. 
\marke{EqTimeProp}\ee
It has no direct physical interpretation, but it is exploited as an 
intermediate step for the definition of the causal propagator 
$U(x,x')$. As a consequence of \Eq{EqTimeProp} and the definition 
\eq{evolPropTime}, the causal propagator $U(x,x')$ satisfies 
\Eq{GreenKleinGordEq}, i.e., it is a Green function of the KG 
equation. 

\subsection{Analysis of the propagator}

We see that the evolution \eq{relatEvolut} of the particle 
is described by the propagator which is expressed 
in the form of a path integral by 
\Eqss{PropagOper}{evolPropTime}{evolPathInt} containing the operator 
\eq{DisplaceOper} of parallel transport. Let us assume that the initial 
state of the particle is localized in a rather small region of 
the spacelike surface $S$ which may be, in a reasonable approximation, 
presented by a single point (the point O in Fig.~\ref{FigEprGrav}) 
and the final state is also concentrated in a small region 
(near the point A$_1$ or A$_2$) of the corresponding 
space-like surface. 

It is known that the main contribution to the path integral 
is given by those paths which are close to geodesic lines (this is 
valid if the initial and final points are not too close to each other). 
Let us assume (this assumption will be discussed later) that the 
uncertainties of the position in the initial and in the final state of 
each particle are small in comparison with the distance between its 
initial and final positions. Then all the paths which substantially 
contribute to the path integral for the given particle are close to 
the same geodesic, namely 
the geodesic line connecting the point of the initial localization O 
of the particle with the point of its final localization (\Aone{} or 
\Atwo). Let us denote the corresponding geodesics as 
$\gamma_1$=O\Aone{} and $\gamma_2$=O\Atwo. Therefore, all paths 
essentially contributing to the evolution of the particle are close to 
$\gamma_1$ for the first particle and close to $\gamma_2$ for the 
second particle. 

The contribution of each path in \Eq{evolPathInt} is 
expressed through the parallel transport along this path. For the 
paths giving the main contribution (close to the geodesic $\gamma_1$ or 
$\gamma_2$) the corresponding parallel transports are close to the 
parallel transport along the corresponding geodesic, $\gamma_1$ for the 
first particle and $\gamma_2$ for the second particle. Therefore, we 
are led to the following conclusion: If the initial spin state 
of the particle is described by a function referred to some 
local frame $n_0$ in the initial point O, then the final spin 
state is approximately described by the same function but 
referred to the local frame (in the final point \Aone{} or \Atwo) 
which is obtained 
from $n_0$ by the parallel transport along the corresponding geodesic 
$\gamma_1$ or $\gamma_2$. Let us denote the resulting local frames 
$n_1$ for the first particle and $n_2$ for the second particle. 

Summing up, the spin state of the first particle in the final point 
\Aone is described by the same function as in the initial point O but 
if the initial spin state refers to the local frame $n_0$ and the 
final spin state refers to the local frame $n_1$ obtained from $n_0$ 
by parallel transport along $\gamma_1$. The same is valid also for the 
second particle, but with $n_2$ and $\gamma_2$ instead of $n_1$ and 
$\gamma_1$.\footnote{This statement may be technically elaborated, 
most conveniently with the help of wave functions defined as functions 
on the fiber bundle of local frames. A spin state of a localized particle 
may be described by a function on the corresponding fiber of local 
frames in the point of localization. If a reference local frame is 
given, then the function on the fiber (representing the spin state) is 
reduced to the function on the Lorentz group. The same function on the 
Lorentz group but referring to the parallelly transported local frame 
will present the final spin state of the particle.}

This evidently justifies the statement made in \Sect{SectPlaus} about 
EPR correlations in a gravitational field. Indeed, the correlation 
between spin states of the two particles is formed at the moment of 
their formation at a single point O. Then the particles propagate to 
different points A$_1$, A$_2$. If this propagation may, in a good 
approximation, be presented as the propagation of the localized states 
along the geodesics $\gamma_1$ and $\gamma_2$, then the spin state of 
each particle will be parallelly transported into the new point of 
localization along the corresponding geodesic. The correlation of the 
initial spin states is described with reference to the same local 
frame $n_0$ (at the point O which is common for both particles). 
Therefore the 
correlation between the final spin states must be described in just 
the same way but with reference to the parallelly translated local 
frames (correspondingly $n_1$ and $n_2$). The relation between these 
local frames is evident: $n_2$ may be obtained from $n_1$ by the parallel 
transport backward along the world line O\Aone=$\gamma_1$ of the first 
particle (which results in $n_0$) and then forward along the world 
line O\Atwo=$\gamma_2$ of the second particle. Finally $n_2$ is the 
parallel transport of $n_1$ along the curve 
\Aone{}O\Atwo=$(\gamma_1)^{-1}\gamma_2$. 

The preceding conclusions were based on the assumption that the 
initial state of each particle is localized in a rather small 
region and the propagation of each particle may, in a good 
appoximation, be presented by a single path coinciding with 
the corresponding geodesic line. Let us discuss this assumption. 

At the first glance it is hardly possible to justify the assumption 
that the evolution of each particle may be accurately enough presented 
by only one geodesic $\gamma_1$ or $\gamma_2$. Indeed, because of the 
uncertainty of the initial linear momentum the localization in the 
final point is poorer than it is in the initial point; the longer 
period of propagation the poorer the localization. The particle may 
be considered localized about the final point \Aone{} or \Atwo{} only 
if the period of propagation is rather short. However for a short 
period of evolution it is impossible to neglect the paths which 
strongly deviate from geodesics. This leads to a contradiction which 
hardly may be resolved. 

In this latter argument we refer to the final localization which is the 
result of the free evolution. This localization is poor because of the 
spreading of the wave packet. However, what we really need is the 
localization of the particles in the measurement here under consideration. 
Let not only the spin projections, but simultaneously with this the 
coordinates of the particles are measured. Then the resulting uncertainty 
of the coordinate will be determined by the precision of the measurement 
of the coordinate. In other words, in the process of the measurement of 
the spin projection the particle may be localized in an arbitrarily 
narrow region about the point \Aone{} or \Atwo{} (these points are 
determined only in the process of this localization during the 
measurement of the spin projections). In this case only those paths in 
the path integral are relevant which lead to this region. We may 
conclude then that only a rather narrow bunch of paths contributes to 
that component of the state which is essential for the analysis. If 
the region of the final localization is narrow enough, then all 
relevant paths are close to the corresponding geodesic ($\gamma_1$ or 
$\gamma_2$), so the assumption we have accepted in the analysis of the 
spin corelation is justified. 

This completes the dynamical proof of the statement which has been 
fromulated in \Sect{SectPlaus} on the basis of plausible arguments. 

It is of course possible that one meets another situation when the 
assumption taken above is invalid. Let the localization of the 
particles (initially or finally or both) be not enough strong to 
justify the approximation of a single path. Then we have to apply a 
more strict approximation taking into account more than one path of 
propagation of each particle. In this case the initial correlation 
between the spin projections of the two particles (necessarily 
resulting in the decay of a scalar particle) is destroyed in the 
course of the propagation. The longer the propagation and the stronger 
the gravitational field, the poorer is the correlation. In general the 
EPR correlation between spin projections does not exist. 

%\newpage
\section{Conclusion}\marke{SectDiscus}

The conclusion drawn from the previous consideration may be formulated 
as follows. What is usually called quantum non-locality (or the 
correlation of the EPR type) is in fact a correlation between the 
results of the measurements of two spatially separated systems. 
However this apparently non-local phenomenon is preceded by 
1)~establishing the correlation between the two systems in a single 
point O (locally) and then 2)~bringing the correlated systems to the 
separated points \Aone{} and \Atwo{} during the evolution 
(propagation). Both operations are local. Thus, the final correlation 
is prepared in the course of the local processes. 

This is made explicit by the above consideration of the EPR effect in 
an external gravitational field. We have seen that the spin states of 
the particles in the points \Aone{} and \Atwo{} are correlated if they 
are referred to such local frames $n_1$ and $n_2$ in these points 
that $n_2$ results from the parallel transport of $n_1$ along the line 
\Aone{}O\Atwo{} consisting of the two geodesics, O\Aone{} and O\Atwo. 

Thus, in the case of the gravitational field (when there is no 
teleparallelism) the very specification of the correlation may be made 
only in terms of the world lines (classical trajectories) of the 
particles (essentially local objects). In general no EPR correlation 
exists in a gravitational field. The reason for this is that different 
local processes (essentially different paths) are coherently 
superposed. 

\vskip 0.5cm
\centerline{\bf ACKNOWLEDGEMENT}

The work was supported in part by the Deutsche Forschungsgemeinschaft 
under grant 436~RUS~17/1/99. 

%\newpage

%\bibliography{bibfilename}
%\bibliographystyle{unsrt}

\end{document}